\documentclass[journal]{IEEEtran}
\ifCLASSINFOpdf
\else
   \usepackage[dvips]{graphicx}
\fi
\usepackage{url}
\usepackage{graphicx}
\usepackage{amsmath,bm}
\usepackage{amssymb}  % 花体字母加粗
\usepackage{cite}        %[1-3]合并参考文献
 \usepackage{epstopdf} %eps 图像
  \usepackage{xcolor} % 字体颜色
  \usepackage{subfigure} % 双图并立
  \usepackage{amsthm}
\begin{document}

\title{Deep Learning based Downlink Channel Prediction for FDD Massive MIMO System}

%\author{Author 1,  Author 2, Author 3, Author 4}
\author{Yuwen Yang,  Feifei Gao, \IEEEmembership{Senior Member, IEEE}, Geoffrey Ye Li, \IEEEmembership{Fellow, IEEE}, 	 Mengnan Jian
\thanks{Manuscript received May 13, 2019; revised July 12, 2019; accepted
August 09, 2019.
This work was supported in part by the National Natural Science Foundation of China under Grant \{61831013,  61771274, 61531011\}, by the Beijing Municipal Natural Science Foundation under Grant 4182030, L182042,  the R\&D Project of the Science, Technology and Innovation Commission of Shenzhen Municipality No. JCYJ20180306170617062, and  is also supported by Zhongxing Telecommunication Equipment (ZTE) fund.
The associate editor coordinating the review of this paper and approving it
for publication was  Van-Dinh Nguyen.\emph{ (Corresponding author: Feifei Gao.)}}
\thanks{Y. Yang and F. Gao  are with  Institute for Artificial Intelligence Tsinghua University
(THUAI), State Key Lab of Intelligent Technologies and Systems, Beijing National Research Center for Information Science and Technology (BNRist), Department of Automation, Tsinghua University, Beijing,
100084, P. R. China. F. Gao  is also with Key Laboratory of Digital TV System of Guangdong Province and Shenzhen City, Research Institute of Tsinghua University in Shenzhen, Shenzhen 518057, P. R. China (email: yyw18@mails.tsinghua.edu.cn, feifeigao@ieee.org).}
\thanks{G.~Y.~Li is with the School of Electrical and Computer Engineering, Georgia Institute of Technology, Atlanta, GA, USA (email: liye@ece.gatech.edu). }
\thanks{M. Jian is with the  Department of Algorithm, Wireless Product R\&D Institute, ZTE Corporation, Beijing 100029, China (email: jian.mengnan@zte.com.cn). }
}

%\markboth{Journal of \LaTeX\ Class Files, Vol. 14, No. 8, April 2019}
\maketitle

\begin{abstract}
In a frequency division duplexing (FDD) massive multiple-input multiple-output (MIMO) system,
the acquisition of  downlink channel state information (CSI) at base station (BS)  is a very challenging task due to the overwhelming overheads required for downlink training and uplink feedback.  In this paper, we  reveal  a deterministic
uplink-to-downlink mapping function
 %for a given communication environment
 when the position-to-channel mapping is bijective.
Motivated by the universal approximation theorem, we then propose a  sparse complex-valued neural network (SCNet) to approximate the uplink-to-downlink mapping function.
Different from   general deep networks that operate in the real domain, the  SCNet is constructed in the complex domain  and is able to learn the complex-valued mapping function by off-line training.
After training, the   SCNet is used to directly  predict  the downlink CSI based on the estimated uplink CSI without the need of either downlink training or  uplink feedback.
Numerical results show that the SCNet achieves better performance than  general deep networks
in terms of    prediction accuracy and exhibits remarkable robustness over complicated wireless channels, demonstrating  its great potential for practical deployments.

\end{abstract}

\begin{IEEEkeywords}
FDD, massive MIMO, downlink CSI prediction, deep learning, complex-valued neural network.
\end{IEEEkeywords}

\IEEEpeerreviewmaketitle
\section{Introduction}\label{secladn}

Massive multiple-input multiple-output (MIMO) has been widely recognized as a promising technique in future wireless communication systems
for its   high spectrum and energy efficiency, high spatial resolution, and large beamforming gains \cite{8354789}.
%\cite{6736761,8354789}.
%\cite{6824752,6736761,6979963,6375940}.\cite{8354789}
To embrace  these benefits, accurate
downlink   channel state inforamtion (CSI) is usually required at both the base station (BS) (for
beamforming, user scheduling, etc) and the user side (for signal detection).
However, the acquisition of downlink CSI  is  a very challenging task
 for frequency division duplexing (FDD) massive MIMO systems due to the prohibitively high overheads
 associated with  downlink training and uplink feedback.

In fact, there are two important observations that can help  reduce the  overheads.
First, the wireless channels between BS and users only have a small angular
spread (AS) as demonstrated in \cite{6328480,zhou2007experimental,hugl2002spatial}.
Due to the small AS and the large dimension of the channels, massive MIMO
channels exhibit  sparsity in the  angular domain.
Secondly,
there exists  angular reciprocity between the uplink and the downlink channels since the uplink and the downlink share the common physical paths \cite{hugl2002spatial}.
Since the acquisition of the uplink CSI is convenient in massive MIMO systems,
many studies have  suggested to  extract partial information of the downlink CSI from the uplink CSI,
thereby reducing the downlink training overhead  or to employ compressive sensing (CS) based algorithms to reduce the overhead of the uplink feedback \cite{8334183,8648511}. %6816089
 For example,
in  \cite{8334183}, the downlink channel covariance matrix (CCM) is first estimated from the uplink CCM
 and then the eigen-beamforming is used to reduce the overhead for the downlink training  when AS is less than $5^\circ$.
 %the eigen-beamforming is adopted to reduce the overhead for DOWNLINK training.
% use the UL channel covariance matrix (CCM) to infer the DOWNLINK CCM, with which
% the overhead for DOWNLINK training can be reduced.
% estimate the DOWNLINK channel covariance matrix (CCM) from the UL CCM,
% and then use the estimated DOWNLINK CCM to
% reduce the DOWNLINK training overhead by exploiting the  DOWNLINK
%channel covariance matrix (CCM) that can be  derived from the UL CCM.
% However, the method in \cite{8334183}  works well only when AS is less than $5^\circ$, which
% restricts its applications in many environments.
In \cite{8648511}, the channels are first
parameterized by  distinct paths, each characterized by
 path delay, angle, and gain. Then, the frequency-independent parameters, i.e.,  path delays and  angles, are extracted from the uplink CSI to help reduce the downlink training.
 Nevertheless, the method in \cite{8648511}
 is applicable as long as the propagation paths are  distinguishable and   the path number is small. %, which can hardly be satisfied in practical  systems.
Besides, several CS-based channel feedback schemes for massive MIMO have been proposed to reduce the feedback overhead but are sensitive to the model errors and suffer from high complexity.

% \cite{6816089}. However, these CS-based methods

Due to its excellent performance and low  complexity \cite{8663966}, deep learning has been
introduced  recently to the wireless physical layer and has achieved superior  performance  over  various topics, such as   channel estimation \cite{8672767},  detection \cite{8052521},
 CSI feedback \cite{8322184}, etc.
 In  \cite{safari2018deep}, a convolutional neural network (CNN) is trained to predict the downlink CSI based on  the  CSI of multiple adjacent uplink subcarriers
 for single-antenna FDD systems.
In \cite{8761962}, a fully-connected neural network (FNN) is trained for uplink/downlink
 channel calibration for  massive MIMO systems.
 % which can also be used for the downlink channel prediction in the FDD massive MIMO systems.
% Specifically, the uplink CSI is separated into  real and imaginary parts  and then fed to  FNN to predict both the  real and imaginary parts  of the downlink CSI.
%Since the FNN can only work in real domain, the uplink CSI is separated into  real and
%imaginary parts  and then fed to the FNN, which potentially losses  the phase information of the input and  destroys the  structure of uplink-to-downlink mapping function.
In this paper,  we propose a sparse complex-valued neural network (SCNet) for the downlink CSI prediction in FDD massive MIMO systems.
Due to the richer representational capacity offered  by complex representations, the SCNet can
further improve the performance of channel prediction.
% To further improve the performance
% of channel prediction, in this paper  we propose a sparse complex-valued neural network (SCNet) to exploit the a richer representational
%capacity offered by complex representations.
 Our  contributions are summarized as follows.
%The basic idea is to training a deep neural network (DNN) to exploit both the sparsity and reciprocity features between the UL and DOWNLINK, thereby predict the  DOWNLINK CSI directly from the UL CSI.
%Our contribution is summarized as follows.
\begin{enumerate}
  \item Inspired by \cite{alrabeiah2019deep}, we  reveal a deterministic uplink-to-downlink
mapping function for a given communication environment when
the position-to-channel mapping is   bijective.
Then, we prove that the uplink-to-downlink mapping function can be   approximated with an arbitrarily small error by a feedforward network.

\item We propose a  SCNet for  downlink CSI prediction in FDD massive MIMO systems, which
 is applicable to complex-valued function approximation with
complex-valued representations. Moreover, sparse network structure is adopted to reduce the complexity and improve the robustness.

  \item Experiment results  demonstrate that  SCNet outperforms  the FNN of \cite{8761962} in terms of prediction accuracy and exhibits remarkable robustness over the number of paths.
\end{enumerate}

\section{System Model}\label{secmodel}
Fig.~\ref{figsbf} illustrates an FDD massive MIMO system,
where the BS is equipped with  $M\gg 1$ antennas in the form of uniform linear array (ULA)\footnote{We adopt the ULA model here for simpler illustration, however,
the proposed approach does not restrict the specifical array shape, and therefore is applicable for an array of arbitrary geometry.} and the user is
equipped with  a single antenna.
 Since   the proposed approach works
 for different users separately, we only need to illustrate for a
 single user.
The  channel between the user and   the BS  is assumed
to consist of $P$ rays and can be expressed as\footnote{Note that we have ignored spatial- and frequency- wideband effects \cite{8354789} as most of the literature. In fact, an accurate mathematical  model
is unnecessary for the   proposed approach since the network can be trained by data from practical systems. }
\cite{8697125},
\begin{equation}\label{equeeed}
\bm h\left(f\right) = \sum\limits_{p = 1}^P {{\alpha _p}{e^{ - j2\pi f\tau_p + j{\phi _p}}}\bm a\left( {{\theta _p}} \right)},
\end{equation}
where $f$ is the carrier frequency, and  $\alpha _p$, $\phi _p$, $\tau_p$ and $\theta _p$ are the  attenuation, phase shift, delay, and direction of arrival (DOA)  of the $p$-th path, respectively. Moreover, $\bm a\left( {{\theta _p}} \right)$ is the the array manifold vector defined as,
\begin{equation}\label{equfed}
\bm a\left( {{\theta_p }} \right)={\left[ {1,{e^{ - j\chi\sin {\theta_p}}}, \cdots ,{e^{ - j\chi\left( {M - 1} \right)\sin {\theta_p }}}} \right]^T},
\end{equation}
where $\chi={{2\pi d f}}/{c }$,
 $d$  is  the antenna spacing, and $c$ is the speed of light.
 According to  \cite{6328480,zhou2007experimental,hugl2002spatial}, the incident AS with mean DOA $\theta$ seen by the BS is   limited in a certain region, i.e., $\theta_p \in \left[\theta-\Delta \theta/2,\theta+\Delta \theta/2\right]$.

Note that $\alpha _p$ depends on (i) the distance between the user and the BS, denoted by $D$,  (ii) the transmitter and receiver antenna gains, (iii) the carrier frequency, and (iv) the scattering environment.
 The phase  $\phi_p$  depends on the scatterer  materials and wave incident/impinging angles at the scatterers.
 The delay $\tau_p$ depends on the distance  travelled by the signal along the $p$-path
  \cite{hugl2002spatial}.

\begin{figure}[!t]%[!hptb] !h意思是忽略美学标准，将照片固定到此位置；不会上下浮动% 支持格式eps, pdf, png, jpg
\centering %使得插入的照片居中显示
\includegraphics[width=80 mm]{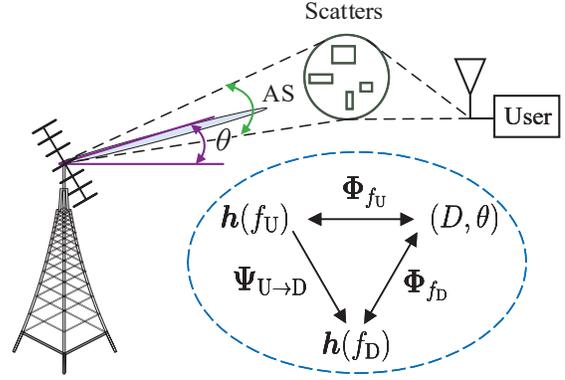}
% 图片标题
\caption{Downlink CSI prediction for FDD Massive MIMO systems.}
\label{figsbfa}       % 给图片一个标签便于交叉引用
\end{figure}
\section{Channel Mapping Formulation}\label{secmap}
%In this section, we first investigate the existence of uplink to downlink mapping,
%First consider the mapping of single user
Denote $\bm h\left(f_{\textrm{U}}\right)$ and
$\bm h\left(f_{\textrm{D}}\right)$ as the uplink and  the downlink channels of the user with
 $f_{\textrm{U}}$ and $f_{\textrm{D}}$ being
the uplink and the  downlink frequencies, respectively.
As indicated in  Eq.~\eqref{equeeed}, $\bm h\left(f_{\textrm{D}}\right)$  cannot be simply obtained from the  $\bm h\left(f_{\textrm{U}}\right)$  for FDD systems.
However, since the downlink and the uplink
experience the same propagation environment with the
common physical paths and the spatial propagation characteristics
of the wireless channels are nearly
unchanged  within certain bandwidth \cite{hugl2002spatial},
there is an  intrinsic relation between the uplink and the downlink CSI.
%we consider whether there exists a  deterministic uplink-to-downlink mapping function.

%the downlink CSI can be obtained by uplink CSI by exploiting these parameter reciprocities.
%
%
%there exist attenuation amplitude, delay and angular reciprocities between the uplink and downlink channels  \cite{hugl2002spatial}. %\cite{hugl2002spatial,vasisht2016eliminating,7174558}.
%Inspired by these facts,
%we consider whether the downlink CSI can be obtained by uplink CSI by exploiting these parameter reciprocities.
%In other words, how can we establish the mapping function from uplink to downlink channels?

 In the following, we first define  an uplink-to-downlink mapping function, following the approach in \cite{alrabeiah2019deep}, and
  prove its existence.
 Then, we   leverage  deep learning  to find the  mapping function.

\subsection{Existence of Uplink to Downlink Mapping}
%We first give the definition and  bijectiveness assumption of the position-to-channel mapping,  based on which the  existence of
% the uplink-to-downlink mapping can be proved.
%investigate the  existence of user to channel mapping, which leads to the uplink to downlink mapping.

Consider the channel model in Eq.~\eqref{equeeed}, where the  channel function $\bm h(f)$ is  completely determined by the parameters  $\alpha_p$, $\phi_p $, $\tau_p$, $P$, $\Delta \theta$, and  $\theta$.  As discussed at the end of Section \ref{secmodel},
$\alpha_p$, $\phi_p $, $\tau_p$, $P$, and $\Delta \theta$  are the functions of the communication environment (including the antenna gains, scatterers, etc.), mean DOA $ \theta$ and  distance $D$.

 \emph{Definition 1:}
 The  position-to-channel mapping ${\bm \Phi _{f}}$  can be written as follows,
\begin{equation}\label{equ}
{\bm \Phi _{f}}:\left\{ {\left(D,\theta\right)} \right\} \to \left\{ \bm h(f )\right\},
\end{equation}
where the sets $\left\{ {\left(D,\theta\right)}  \right\}$ and $\left\{ \bm h(f )\right\}$ are the domain and codomain of the mapping $\bm \Phi _{f}$, respectively.

 Then, we adopt the following assumption  for further analysis.

 \emph{Assumption 1}\cite{alrabeiah2019deep}{:} The position-to-channel mapping function, ${\bm \Phi _{f}}:\left\{ {\left(D,\theta\right)} \right\} \to \left\{ \bm h(f )\right\}$,
is bijective.

 The \emph{Assumption 1} means that every user position  has a unique channel function $\bm h(f)$, and vice versa. Although it cannot be proved analytically,
 the  probability that
$\bm \Phi _{f}$ is bijective is actually very high in
practical wireless communication scenarios, and  approaches 1 as the number of antennas at the BS increases \cite{alrabeiah2019deep}. Therefore,
it is reasonable to adopt  \emph{Assumption 1} in   massive MIMO systems.
%\footnote{In fact, the \emph{Assumption 1} has been widely adopted in the wireless positioning and fingerprinting literature \cite{8292280,8554268,8509634}.}.

Under \emph{Assumption 1}, the channel-to-position mapping, i.e.,  the inverse mapping of ${\bm \Phi _{f}}$,
 exists, which can be written as follows:
\begin{equation}\label{equinv}
{\bm \Phi _{f}^{-1}}:\left\{ \bm h(f )\right\} \to \left\{ {\left(D,\theta\right)} \right\}.
\end{equation}

Next, we investigate the  existence of the uplink-to-downlink mapping, as given in
\emph{Proposition 1}.

\emph{Proposition 1:}
With \emph{Assumption 1},  the uplink-to-downlink mapping exists for a given communication environment, and can be written as follows,
\begin{eqnarray}
\bm \Psi _{{\textrm{U}} \to {\textrm{D}}} \!\!\!\! &=& \!\!\!\! {\bm \Phi _{f_{\textrm{D}}}}\circ{\bm \Phi _{f_{\textrm{U}}}^{-1}} : \left\{ \bm h(f_{\textrm{U}} )\right\} \to \left\{ \bm h(f_{\textrm{D}} )\right\},
\end{eqnarray}
%\begin{eqnarray}
%% \nonumber to remove numbering (before each equation)
%\bm \Psi _{{\textrm{U}} \to {\textrm{D}}} \!\!\!\! &=& \!\!\!\! {\bm \Phi _{f_{\textrm{D}}}}\circ{\bm \Phi _{f_{\textrm{U}}}^{-1}} \nonumber \\
%  \ \!\!\!\!&:& \!\!\!\! \left\{ \bm h_{k}(f_{\textrm{U}} )\right\} \to \left\{ \bm h_{k}(f_{\textrm{D}} )\right\},
%\end{eqnarray}
where  ${\bm \Phi _{f_{\textrm{D}}}}\circ{\bm \Phi _{f_{\textrm{U}}}^{-1}} $ represents the
composite mapping related to ${\bm \Phi _{f_{\textrm{D}}}}$ and ${\bm \Phi _{f_{\textrm{U}}}^{-1}} $.
\begin{proof}
From the \emph{Definition 1}, we have the mappings
$  {\bm \Phi _{f_{\textrm{D}}}} : \left\{ {\left(D,\theta\right)} \right\} \to \left\{ \bm h(f_{\textrm{D}} )\right\}$ and ${\bm \Phi _{f_{\textrm{U}}}} :\left\{ {\left(D,\theta\right)} \right\} \to \left\{ \bm h(f_{\textrm{U}} )\right\}$
exist in position candidate set $\left\{ {\left(D,\theta\right)} \right\}$.
Under \emph{Assumption 1}, the mapping ${\bm \Phi _{f_{\textrm{U}}}^{-1}} $ exists with  its codomain
equal to the domain of $ {\bm \Phi _{f_{\textrm{D}}}}$. Therefore, the composite mapping
${\bm \Phi _{f_{\textrm{D}}}}\circ{\bm \Phi _{f_{\textrm{U}}}^{-1}} $ exists for any possible  position
${\left(D,\theta\right)}$.
%The proof is illustrated in Fig.~\ref{figsbfa}.
\end{proof}

A more general proposition can be found in \cite{alrabeiah2019deep}.

%One underlying assumption for \emph{Proposition 1} is that  users are not in high-speed scenarios, otherwise the communication environment of  uplink transmission  is different with that of downlink.
%Uplink-to-downlink mapping in high-speed scenarios will be left as future work.

%the uplink and downlink transmission are within one  coherence time period. For fast time-varying scenarios, the user position in uplink transmission is typically
%High-Speed Scenarios
%high mobility
%Recalling the channel model in Eq.~\eqref{equfed}, in the $n$-th coherence time period $T_n$, the $k$-th user can be uniquely identified  by  $\left\{T_n,D_k,\mathbb{A}_k\right\}$. For a given communication   environment (including the antenna gains and positions, scatterers, etc.),
%there exists a deterministic mapping function from the $k$-th user  to the channel, since
%$\left| {{\alpha _k}\left( \theta  \right)} \right|$, $\phi \left( \theta  \right)$ and $\tau_k \left( \theta  \right)$ are the functions of scatterer environment and carrier frequency.
%\begin{eqnarray}
%% \nonumber to remove numbering (before each equation)
%  {\bm \Phi _{k,f_{\textrm{D},k}}} {\kern -6pt}&:& {\kern -6pt} \left\{ {\left(D_k,\mathbb{A}_k\right)} \right\} \to \left\{ \bm h_{k}(f_{\textrm{D},k} )\right\}; \\
% {\bm \Phi _{k,f_{\textrm{U},k}}} {\kern -6pt}&:& {\kern -6pt} \left\{ {\left(D_k,\mathbb{A}_k\right)} \right\} \to \left\{ \bm h_{k}(f_{\textrm{U},k} )\right\}.
%\end{eqnarray}

\subsection{Deep Learning for Uplink-to-Downlink Mapping}
\emph{Proposition 1} proves the existence of the uplink-to-downlink mapping function. However, the function cannot be depicted by known mathematical models, which motivates us to resort to deep learning algorithms.
Based on the universal approximation theorem \cite{hornikmultilayer}, we obtain  \emph{Theorem~1} as following.

\emph{Theorem 1:}
For  any given small error  $\varepsilon>0$, there always exists a positive
constant $N$ large enough such that
\begin{align}
\mathop {\sup }\limits_{\bm x \in \mathbb{H}} \left\| {\textrm{NET}_{N}\left( \bm x, \bm \Omega \right)-\bm \Psi _{{\textrm{U}} \to {\textrm{D}}} \left( \bm x \right)   } \right\| & \le \varepsilon, \ \mathbb{H}=\left\{ { \bm h(f_{\textrm{U}} ) } \right\},
\end{align}
where $\textrm{NET}_{N}\left( \bm x, \bm \Omega \right)$ is the output of a three-layer feedforward network with $\bm x$, $\bm \Omega$ and $N$
 denoting the input data, network parameters, and the number of hidden units,  respectively.
\begin{proof}
(i) Since  $  \bm h(f_{\textrm{U}} ) $ is bounded and closed, $ \mathbb{H} $ is a  compact set;
 (ii) Since ${\bm \Phi _{f_{\textrm{D}}}}$  and ${\bm \Phi _{f_{\textrm{U}}}^{-1}}$ are  continuous mapping and the composition of continuous mappings is still a continuous mapping, we know that for $\forall \bm x \in \mathbb{H}$ such that  $\bm \Psi _{{\textrm{U}} \to {\textrm{D}}} \left( \bm x \right)   $ is a continuous function.
Based on (i), (ii), and  the universal approximation theorem \cite[Theorem 1]{hornikmultilayer},
  \emph{Theorem~1} is proved.
\end{proof}

According to \emph{Theorem 1},  the uplink-to-downlink mapping function can be approximated with an
arbitrarily small error  by a feedforward network with a single hidden layer.
Thus,  we can train a network to predict the downlink CSI from the uplink CSI
and can significantly
reduce the overhead required for downlink training and
uplink feedback at the cost of off-line training.

\begin{figure}[!t]%[!hptb] !h意思是忽略美学标准，将照片固定到此位置；不会上下浮动% 支持格式eps, pdf, png, jpg
\centering %使得插入的照片居中显示
\includegraphics[width=85 mm]{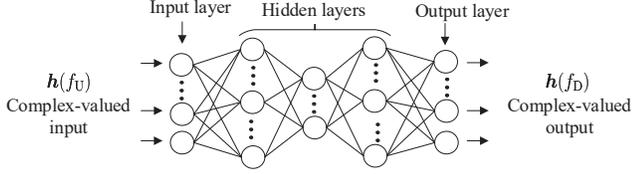}
% 图片标题
\caption{The SCNet  architecture.}
\label{figsbf}       % 给图片一个标签便于交叉引用
\end{figure}

\section{SCNet based Downlink CSI Prediction}
In this section, we will first introduce the architecture of  the SCNet. Then, we discuss how to
train and deploy it in massive MIMO systems.
\subsection{SCNet Architecture}
Although it has been proven in \emph{Theorem~1} that a three-layer network is able to predict the downlink CSI, we propose the SCNet instead of the three-layer network for  practical considerations as follows:
(i) A  deep network with an appropriate number of layers learns better than a three-layer network;
(ii) A spare network can reduce the network parameters, and therefore is easier to train and is more robust;
(iii) Compared with the real-valued networks,
the complex ones have  richer representational capacities and therefore are  more powerful in learning complex-valued functions \cite{trabelsi2017deep}.

As shown in Fig.~\ref{figsbf}, the input of the SCNet is the uplink CSI  $  \bm h(f_{\textrm{U}} ) $.
The output of  the SCNet is a cascade of nonlinear transformation of $  \bm h(f_{\textrm{U}} ) $, i.e.,
\begin{equation}\label{equnn}
\hat{ \bm h}(f_{\textrm{D}} ) = \textrm{NET}\left(  \bm h(f_{\textrm{U}} ), \bm\Omega \right)=\bm f^{(L-1)}\left(\cdots \bm f^{(1)}\left( \bm h\left((f_{\textrm{U}}\right)\right)\right),
\end{equation}
where $L$ is the number of layers and $\bm\Omega\buildrel \Delta \over =\left\{\bm W^{(l)},\bm b^{(l)}\right\}_{l=1}^{L-1}$  is the network parameters to be trained.
Moreover, $\bm f^{(l)}$  is the nonlinear transformation function of the  $l$-th layer and can be written as,
\begin{align}
\begin{array}{l}%array视作矩阵，表示矩阵表格只有一列，并向左对齐
\bm f^{(l)}\left( \bm x \right) = \left\{
\begin{array}{ll} %{ll}表示两列向左对齐{rr}向右。&其实是列的分隔符，若要对等号对齐，单独开辟一列给等号，形式上表现为&=&
\bm g\left(\bm W^{(l)}\bm x +\bm b^{(l)}\right),\ & 1\le l<L-1;\\
\bm W^{(l)}\bm x +\bm b^{(l)}, \ & l=L-1,
\end{array} \right.
\end{array}
\end{align}
where $ \bm g $ is the activation function and is given by
\begin{equation}\label{eqeywu}
 \bm g (\bm z)= \max{\left\{{\Re[\bm z],\bm 0}\right\}}+j\max{\left\{{\Im[\bm z],\bm 0}\right\}}
\end{equation}
with  $\Re[\cdot]$ and $\Im[\cdot]$  being  the real and imaginary parts of the vectors, respectively.

We set the number of neurons in the middle hidden layer to be  much fewer than that in the output layer, which forces  the SCNet to compress the representation of the input.
We would like to emphasize that the compression task
would be very difficult if the elements of input $\bm x$  are independent of each other.
However, since there exists the sparse structure\footnote{ Since  AS is narrow,  the massive MIMO channels exhibit sparsity
in the  angular domain. See more details in \cite{8354789}.} in the uplink channel $  \bm h(f_{\textrm{U}} ) $,  the SCNet is able to discover the intrinsic  sparsity of $  \bm h(f_{\textrm{U}} ) $ in massive MIMO systems.
As a result, the SCNet can not only reduce the redundancy of network parameters but also become more functional and robust \cite{ng2011sparse}.
\begin{figure*}[!t]
\centering     %%% not \center
\subfigure[NMSE performance versus AS $\Delta \theta $]{\label{figsbfas}\includegraphics[width=85mm]{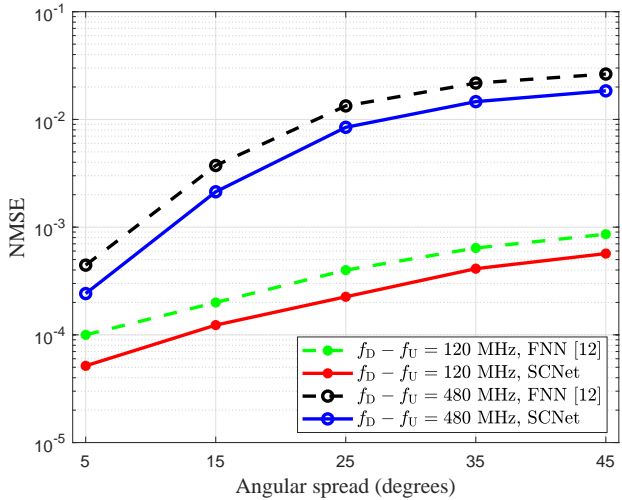}}
\subfigure[NMSE performance  versus  frequency difference $f_\textrm{D}-f_\textrm{U}$ ]{
\label{figsbfdf}\includegraphics[width=85mm]{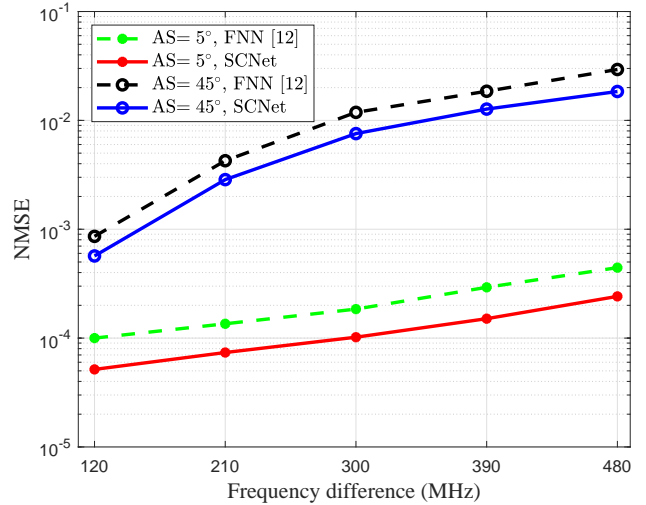}}
\caption{The NMSE performance of the SCNet and the FNN based downlink CSI predictors  versus  AS (a) and the frequency difference
$f_\textrm{D}-f_\textrm{U}$ (b).}
\label{figtrackingfggh}
\end{figure*}
\subsection{Training and Deployment}
%In the off-line training stage,  BS collects both the downlink CSI, which is fed back from the user, and the uplink CSI, which is estimated at BS, as training samples  to train the SCNet.

The proposed downlink CSI prediction has two stages, i.e., the  off-line training and the on-line deployment stages.
In the off-line training stage, the BS collects both the downlink and the uplink CSI as training samples  to train the SCNet. Specifically, during a coherence time period,  the downlink CSI is first estimated at the user side by downlink training and then fed  back to the BS. The uplink CSI is estimated at the BS by uplink training.
 The SCNet is trained to minimize the difference between the output $\hat{ \bm h}(f_{\textrm{D}} )$ and the supervise label ${ \bm h}(f_{\textrm{D}} )$.
The loss function is
\begin{equation}\label{equloss}
\textrm{Loss}\left(\bm \Omega\right) =\frac{1}{{V{N_h}}}\sum\limits_{v= 0}^{ V-1}\left\|\hat{ \bm h}(f_{\textrm{D}} )^{(v)}-{ \bm h}(f_{\textrm{D}} )^{(v)} \right\|_{2}^{2},
\end{equation}
where $V$ is the batch size\footnote{Batch size is the number of samples in one training batch.}, the superscript $(v)$ denotes
the index of the $v$-th  training sample,  ${\left\|{\cdot}\right\|}_{2}$ denotes the  $\ell_2$ norm, and $N_h$ is  the length of the vector  ${ \bm h}(f_{\textrm{D}} )$.
The loss function $\textrm{Loss}\left(\bm \Omega\right) $  is  minimized by the complex designed  adaptive moment estimation (ADAM)  algorithm  \cite{trabelsi2017deep} until
the SCNet converges.

In the deployment stage, the parameters of the SCNet are fixed.  The SCNet directly generates the prediction of the downlink CSI
$\hat{ \bm h}(f_{\textrm{D}} )$ based on the uplink CSI ${ \bm h}(f_{\textrm{U}} )$.
%The UL CSI $\bm h_{\textrm{UL}}$ is reshaped following Eq.~\eqref{equreshape} and  fed into the SSnNet. Then, the predicted downlink CSI, denoted by $\hat{\bm h}_{\textrm{downlink}}$, can be obtained by reshaping the output of the SSnNet following Eq.~\eqref{equreshape2}.
% 4. deploy  beamforming  equation + detection BER
% statistics from the training stage to test the robustness of the
%downlink-based estimation algorithm.
%5. complexity

\subsection{Complexity Analysis}
Denote $n_l$ as the number of  neurons in the $l$-th layer.
The required number of floating point operations (FLOPs)  is used as the metric of complexity.
For real-valued network,  the total number  of FLOPs required is $\sum_{l=1}^{L-1} n_{l-1}n_{l}$.
However, as a complex multiplication is 4 times of its real
counterpart,  the total number of FLOPs required in  the SCNet is $4\sum_{l=1}^{L-1} n_{l-1}n_{l}$.
Nevertheless,
it should be noted that in a real-valued network, the input complex data are typically separated to real and imaginary parts and then fed to the network. Therefore, the  size of  the real-valued network  is lager than  that of a complex-valued network.

\section{Simulation Results}
Unless otherwise specified, the system parameters are set as follows:
 the BS is  equipped with 128  antennas; the uplink frequency follows the 3GPP R15 standard, i.e.,
$f_\textrm{U}=2.5$ GHz. In the simulation of Section \ref{nmse}, the  attenuation of each path follows Rayleigh distribution. The phase and delay of each path follow uniform distribution over $[-\pi,\pi)$ and $[0,10^{-4}]s$. The number of paths is 200 in both the training and deployment stages. While in the simulation of Section \ref{robus}, the parameters of each path are generated according to the ray-tracing simulator \cite{timmurphy}. The number of paths is 200 in  the training  stage and  varies  in the deployment stage.

 %Graphical Processing Units

 An FNN in  \cite{8761962}  is originally designed  for uplink/downlink
 channel calibration for  massive MIMO systems, which
 can also be used for the downlink channel prediction in the FDD massive MIMO systems.
Therefore,  the FNN  is used as a benchmark in this paper. % with its architecture provided by \cite{8761962}.
Keras 2.2.0 is employed as the deep learning framework for both the SCNet and  the FNN.
We choose the number of neurons in the hidden layer as (128, 64,  128) by trails and adjustments.
%\footnote{The  hyper-parameters  provided here may not be optimal but are fairly effective. The performance of the SSnNet can be further improved by fine tuning the  hyper-parameters.}
The initial learning rate of the ADAM algorithm is 0.001. The batch size is 128.
The  parameters of the  SCNet are initialized as
  complex distribution  with normalized variance\footnote{The  weights of  neurons in the $l$-th layer are initialized as complex normal   variables with  variance   $1/n_l$.}.
  The uplink CSI fed to the SCNet is estimated by the minimum mean-squared-error (MMSE) algorithm when the signal-to-noise ratio (SNR) is 25 dB.
 The network is trained for each AS degree and each downlink frequency separately.
The number of training samples is 102,400, and the number of epochs is 400.
\subsection{Prediction Accuracy versus AS and Frequency Difference}\label{nmse}
%\begin{figure}[!t]%[!hptb] !h意思是忽略美学标准，将照片固定到此位置；不会上下浮动% 支持格式eps, pdf, png, jpg
%\centering %使得插入的照片居中显示
%\includegraphics[width=88 mm]{as.eps}
%% 图片标题
%\caption{The NMSE performance of the SSnNet based downlink CSI prediction versus AS.}
%\label{figsbfas}       % 给图片一个标签便于交叉引用
%\end{figure}
%\begin{figure}[!t]%[!hptb] !h意思是忽略美学标准，将照片固定到此位置；不会上下浮动% 支持格式eps, pdf, png, jpg
%\centering %使得插入的照片居中显示
%\includegraphics[width=89 mm]{df.eps}
%% 图片标题
%\caption{The NMSE performance of the SSnNet based downlink CSI prediction versus AS.}
%\label{figsbfdf}       % 给图片一个标签便于交叉引用
%\end{figure}

Normalized MSE (NMSE) is used to measure the prediction accuracy, which is defined as
\begin{equation}\label{equmse}
\textrm{NMSE} =E\left[\left\|{\bm h}_{\textrm{D}}-\hat{\bm h}_{\textrm{D}}\right\|_{2}^{2}/\left\|{\bm h}_{\textrm{D}}\right\|_{2}^{2}\right],
\end{equation}
where $E\left[\cdot\right]$ represents the expectation operation.

Fig.~\ref{figtrackingfggh}  depicts the NMSE performance of  the SCNet and the FNN  based downlink CSI predictors versus AS $\Delta \theta $ and frequency difference $f_\textrm{D}-f_\textrm{U}$, respectively.
 Fig.~\ref{figsbfas} shows that the NMSE performance of both  the SCNet and the FNN degrades as  AS increases while the slope of the NMSE curve decreases as  AS increases. This is because that as  AS increases, the sparsity of channels in the angular domain  decreases,
 and thus it is harder for the networks to learn the structure  of the channels and to accurately predict the downlink CSI.
The networks are less sensitive to  AS  in the wider AS case\footnote{More specifically,
   AS increases by 100\% from $5^\circ$
 to $10^\circ$
  while it increases  by 25\% from $20^\circ$
 to $25^\circ$. The decreasing proportion renders  the SCNet less sensitive to  AS in wide AS case. }, which  accounts for
 the decrease of the  slope in the wide AS case.
Fig.~\ref{figsbfdf} shows that the NMSE performance of both  the SCNet and the FNN degrades as the frequency difference increases.
 This is because the
correlation of CSI between the uplink and the downlink tends to vanish as the frequency difference  increases.
%This is because the angular reciprocity between the UL and downlink tends to vanish as the frequency difference  increases.
As shown in Fig.~\ref{figtrackingfggh},
the proposed the SCNet outperforms  the FNN in all scenarios, which validates
that   the SCNet can benefit from the
rich representational capacity offered by complex representations.

\begin{figure}[!t]%[!hptb] !h意思是忽略美学标准，将照片固定到此位置；不会上下浮动% 支持格式eps, pdf, png, jpg
\centering %使得插入的照片居中显示
\includegraphics[width=85 mm]{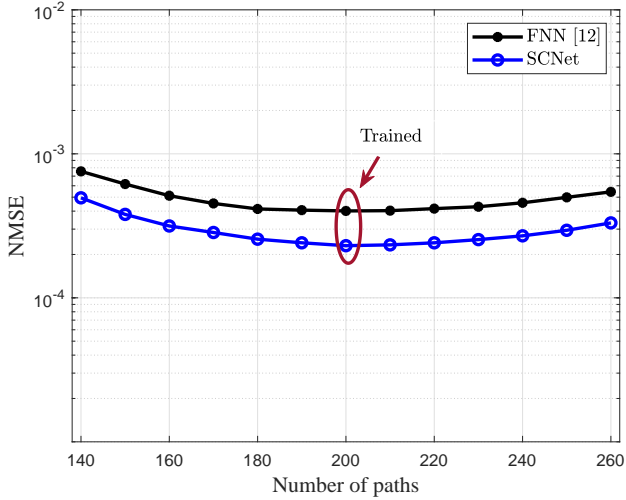}
% 图片标题
\caption{The NMSE performance of  the SCNet and the FNN \cite{8761962} based downlink CSI predictors when mismatches occur between the training and deployment stages.}
\label{figsbfrobust}       % 给图片一个标签便于交叉引用
\end{figure}

%Bit-error rate (BER) is used to measure the detection accuracy and the modulation mode is
%quadrature phase-shift keying (QPSK). As shown in Fig.~\ref{figsber}, the BERs are all lower than $10^{-6}$ for all frequency differences when SNR is 15 dB, which demonstrates the excellent performance of the SSnNet based approach in downlink detection.

\subsection{Robustness Analysis}\label{robus}

In Sections \ref{nmse}, the channels are generated based on Eq.~\eqref{equeeed}  with  the same statistics. However, channels in real-world may be more complicated and the statistics mismatches between the training and deployment stages
are also inevitable.
To test the robustness of both  the SCNet and the FNN, data generated from Wireless InSite \cite{timmurphy} under different scenarios are used to train and test.
As shown in Fig.~\ref{figsbfrobust}, the number of paths
in the training stage is 200 while it varies in the deployment stages. The results show that the variations on statistics of channel  degrade the performance,
but  the SCNet and the FNN still exhibit remarkable prediction accuracy, which validates the excellent generalization ability of  deep neural networks.  %Furthermore, the networks are more robust when the number of paths is larger. This is because that a lager number of paths leads to a more stable statistics property of the CSI, which improves the performance of the networks.

\section{Conclusion}
In this paper, we
   revealed the existence of a deterministic uplink-to-downlink mapping function for a given communication environment.
Then, we proposed  the SCNet  for the downlink CSI prediction in FDD massive MIMO systems.
%In this paper, we have proposed the the SCNet based framework for downlink CSI prediction in FDD massive MIMO systems.
%After training,  the SCNet
%can directly predict the downlink CSI based on observed uplink CSI
%without overhead required for downlink training and uplink feedback.
Simulation results have demonstrated that
  the SCNet performs better than the existing network
 in terms of prediction accuracy.
Furthermore, the remarkable robustness of   the SCNet with respect to the statistic characteristics of wireless channels  has shown its great potential  in real-world applications.
%In addition, a few pilots can be used at users to enhance the accuracy of downlink CSI estimation.

%\bibliographystyle{IEEEbib}
%\bibliography{References}

\end{document}